\documentclass[twocolumn,aps,showpacs,floatfix,prc]{revtex4}
\usepackage[dvips]{epsfig}
\begin{document}
%\draft

\title{ {\Large $\gamma$} induced multiparticle emissions of medium mass nuclei at intermediate energies }

\author{ Tapan Mukhopadhyay$^1$ and D. N. Basu$^2$ }
\affiliation{Variable  Energy  Cyclotron  Centre, 1/AF Bidhan Nagar, Kolkata 700 064, India }

\email[E-mail 1: ]{tkm@veccal.ernet.in}
\email[E-mail 2: ]{dnb@veccal.ernet.in}

\date{\today }

\begin{abstract}

    A comprehensive analysis of multiparticle emissions following photon induced reactions at intermediate energies is provided. Photon induced reaction is described in the energy range of $\sim 30-140$ MeV with an approach based on the quasideuteron nuclear photoabsorption model followed by the process of competition between light particle evaporation and fission for the excited nucleus. The evaporation-fission process of the compound nucleus is simulated in a Monte-Carlo framework. The study shows almost no fission events for the medium mass nuclei and reproduces the available experimental data of photonuclear reaction cross sections satisfactorily at energies $\sim 30-140$ MeV.
\end{abstract}

\maketitle

\noindent
%\section{Introduction}
\label{section1}

    With energetic incident photon it is possible to induce nuclear reactions including fission in most elements. Photonuclear reactions in different energy ranges, like giant dipole resonance (GDR) and quasideuteron (QD) energy regions, have been studied in the past as they provide a wide range of information either on the initial nuclear excitation mechanism or on characteristics of the compound nucleus decay channels. In fact photoneutron cross sections in the GDR energy region have been compiled \cite{Di88} for most of the nuclei in the periodic table, while measurements in the QD region were mainly focussed on heavy nuclei, thus leading to a lack of study on intermediate nuclei. For heavier nuclei, particularly for actinides and preactinides, with high enough energy of incident photons, the dominant reaction mechanism is fission where photofission cross sections are quite large \cite{Mu07}. With decreasing  fissility ($Z^2/A$) of the target nucleus and energy of incident photon, fission probability decreases, whereas other reaction channels such as ($\gamma,xn$) and ($\gamma,xnyp$) type become more important. Thus it becomes important to study the photonuclear reactions in the intermediate energy range for medium mass elements. Investigations of such multiparticle emissions following photon induced reactions in the intermediate energy region in medium mass elements may reveal important features of the nature of the photoabsorption mechanism as well as the decay channel characteristics.

    The aim of the present work is to investigate photonuclear reactions of the type ($\gamma,xn$) and ($\gamma,xnyp$). The present study is restricted to the QD region which is above the GDR or the isovector giant quadrupole resonance (IVGQR) regions and is below the pion threshold. The different  photonuclear reactions for V, Zn, Sn, Ce, Sm, Yb, Ta, Au, Pb and Bi are consistently described as a two step process. In the rapid stage of a photonuclear reaction, the incoming $\gamma$ is assumed to be absorbed by a neutron-proton [n-p] pair inside the nucleus. The rapid stage is then followed by a subsequent de-excitation of the compound nucleus (CN) via evaporation. Such a statistical decay of the CN is the slow stage of a photonuclear reaction. The quantitative description of the process is based on the liquid drop model (LDM) for nuclear fission by Bohr and Wheeler \cite{Bo39} and the statistical model of nuclear evaporation developed by Weisskopf \cite{We37}. For this slow stage, an evaporation Monte-Carlo routine, based on the Weisskopf statistical theory, is used to address the de-excitation of the CN in terms of the competition between particle evaporation and nuclear fission.

\noindent
%\section{The photoabsorption mechanism}
\label{section2}

    The dominant mechanism for nuclear photoabsorption at intermediate energies is described by the QD model \cite{Le51} which is employed to access \cite{Le79} the total photoabsorption cross section in nuclei. It is based on the assumption that the incident photon is absorbed by a correlated n-p pair inside the nucleus, leaving the remaining nucleons as spectators. Such an assumption is enforced when wavelength of the incident photon is small compared to nuclear dimensions but not too small, where the wavelength becomes smaller than the average internucleon spacing in the nucleus. The total nuclear photoabsorption cross section $\sigma_a^T$ is then proportional to the available number of n-p pairs inside the nucleus and also to the free deuteron photodisintegration cross section $\sigma_d(E_{\gamma})$ :

\begin{equation}
%\vspace{-0.14cm}
 \sigma_a^T(E_{\gamma}) = \frac{L}{A} NZ \sigma_d(E_{\gamma}) e^{-D/E_\gamma}
\label{seqn1}
%\vspace{-0.14cm}
\end{equation}
\noindent
where $N$, $Z$ and $A$ are the neutron, proton and mass numbers respectively, $\frac{L}{A}$ factor represents the fraction of correlated n-p pairs and the function $e^{-D/E_\gamma}$ accounts for the reduction of the n-p phase space due to the Pauli exclusion principle. A systematic study of total nuclear photoabsorption cross section data in the intermediate energy range shows that $D = 0.72 A ^{0.81}$ MeV \cite{Te89}. The free deuteron photodisintegration cross section \cite{Wu77} is given by

\begin{equation}
  \sigma_d(E_\gamma) = \frac{61.2 ~ [E_\gamma - B]^{3/2}}{E_\gamma^3} ~ [{\rm mb}]
\label{seqn2}
\end{equation}   
\noindent
where $B~(=2.224$ MeV) is the binding energy of the deuteron. The QD model \cite{Le51,Le79} of nuclear photoabsorption is used together with modern rms radius data to obtain Levinger's constant $L=6.8-11.2A^{-2/3}+5.7 A^{-4/3}$ of
nuclei throughout the Periodic Table and is in good agreement \cite{Ta92} with those obtained from the experimentally measured $\sigma_a^T$ values.

    At the QD energy range, as a consequence of the primary photointeraction, $\gamma$+(n+p) $\rightarrow$ n*+p*, in most of the cases excited compound nucleus is formed with the same composition as the target nucleus where both neutron and proton are retained inside the nucleus and the probabilities that either the neutron or the proton or both escape from the nucleus are extremely low. Hence, the recoiling nucleus can be viewed as a compound nucleus having the same composition as the target nucleus but with excitation energy $E^*=m_0 c^2 [(1+2E_\gamma/m_0 c^2)^{1/2}-1]$ \cite{Ba08}, where $E_\gamma$ is the incident photon enegy and $m_0$ is the rest mass of the nucleus before photon absorption. This excited compound nucleus then undergoes successive evaporation of neutrons, protons and light particles or fission. Hence the photonuclear reaction cross section $\sigma_r$ is a product of the nuclear photoabsorption cross section $\sigma_a^T$ and the statistical decay probability $p$ for a reaction channel and is, therefore, given by $\sigma_r(E_\gamma)=\sigma_a^T(E_\gamma)p$. The decay probability $p$ is basically the branching ratio $\Gamma_r/\Gamma$ where $\Gamma_r$ and $\Gamma$ are the partial and total reaction widths respectively.

    The evaporation stage is calculated by applying the statistical theory \cite{We37}. The decay of the CN takes into account all accessible channels, with the related branchings calculated in terms of the nuclear level densities of the daughter nuclei. The basic steps are the calculations of the relative probabilities between the competing channels (particle evaporation and nuclear fission). The probability of fission relative to neutron emission is calculated using Vandenbosch-Huizenga's equation \cite{Hu73}, given by

\begin{equation}
 \frac{\Gamma_f}{\Gamma_n} = \frac{K_0 a_n[2(a_f E_f^*)^\frac{1}{2}-1]}{4A^\frac{2}{3}a_f E_n^*} \exp{{\big [}2 [(a_f E_f^*)^\frac{1}{2} - (a_n E_n^*)^\frac{1}{2}] {\big ]}}
\label{seqn3}
\end{equation}
\noindent
where $E_n^*=E^*-B_n$ and $E_f^*=E^*-B_f$ are the nuclear excitation energies after the emission of a neutron and after fission, respectively, where $B_n$ is the binding energy of the emitted neutron. $\Gamma_n$ and $\Gamma_f$ are the partial widths for the decay of the excited compound nucleus via neutron emission and fission, respectively, and the parameters $a_n$ and $a_f$ are the level density parameters for the neutron emission and the fission, respectively and $K_0=\hbar^2/2mr^2_0$ where $m$ and $r_0$ are the neutron mass and radius parameter respectively. The emission probability of particle $k$ relative to neutron emission is calculated according to the Weisskopf's statistical model \cite{We37}

\begin{figure}[htbp]
\vspace{-0.0cm}
\eject\centerline{\epsfig{file=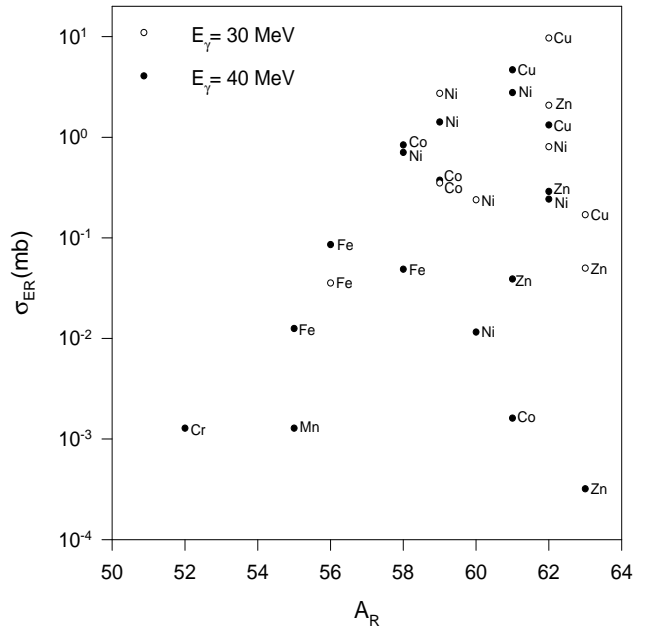,height=8.28cm,width=8.28cm}}
\caption
{The plots of cross sections $\sigma_{ER}$ as a function of mass number $A_R$ of the evaporation residues resulting from ($\gamma,xn$) and ($\gamma,xnyp$) types of reactions for $^{64}$Zn at $E_\gamma = 30, 40$ MeV.}
\label{fig1}
\vspace{-0.0cm}
\end{figure}
\noindent

\begin{equation}
 \frac{\Gamma_k}{\Gamma_n} = {\Big (}\frac{\gamma_k}{\gamma_n}{\Big )} {\Big (}\frac{E_k^*}{E_n^*}{\Big )}  {\Big (}\frac{a_k}{a_n}{\Big )} \exp{{\big [}2 [(a_k E_k^*)^\frac{1}{2} - (a_n E_n^*)^\frac{1}{2}] {\big ]}}
\label{seqn4}
\end{equation}

\noindent
where $a_k$ is the level density parameter for the emission of the particle $k$, $\gamma_k/\gamma_n =1$ for $k=p$, 2 for $k=^4$He, 1 for $k=^2$H, 3 for $k=^3$H and 2 for $k=~^3$He. $E_k^* = E^* - (B_k + V_k)$ is the nuclear excitation energy after the emission of particle $k$ \cite{Le50}. $B_k$ are the binding energies of the emitted particles and $V_k$ are the Coulomb potentials. The evaporation-fission competition of the compound nucleus is then described in the framework of a projection angular momentum coupled \cite{Ha52} Monte-Carlo routine \cite{Pa84}. Any particular reaction channel $r$ is then defined as the formation of the compound nucleus via photoabsorption and its decay via particle emission or fission. Thus, fission is considered as a decay mode. The photonuclear reaction cross sections $\sigma_r$ are calculated using the equation $\sigma_r = \sigma_a^T n_r/N$ where $n_r$ is the number of events in a particular reaction channel $r$ and $N$ is total number of events that is the number of the incident photons.

\begin{table*}[htbp]
\vspace{-0.0cm}
\caption{\label{tab:table1} Photofission cross section and three largest photonuclear reaction cross sections in ($\gamma,xnyp$) decay channels. The number of evaporated neutrons $x$ and protons $y$ are provided as $(x,y)$ adjacent to each reaction cross sections.}
\begin{center}
\begin{tabular}{cccccccccc}
\hline
\hline
Target&$Z^2/A$&Calculated&$E_\gamma$ [MeV]&$E_\gamma$ [MeV]&$E_\gamma$ [MeV]&$E_\gamma$ [MeV]&$E_\gamma$ [MeV]&$E_\gamma$ [MeV]&$E_\gamma$ [MeV]      \\ 
nuclei&   &quantity [mb]&30 &40 &60 &80 &100 &120 &140  \\
\hline 
\hline
$^{51}$V&10.37&$\sigma_a^T$&14.11 &10.92 &7.07 &5.01 & 3.78 & 2.97 &2.42 \\
       &&$\sigma_r(\gamma,xnyp)$&10.76 (2,0)&4.30 (3,0)&3.46 (3,1)&1.80 (4,1)&0.95 (4,2)&0.55 (5,2)&0.31 (5,3) \\
                 &&&                      2.72 (1,1)&2.36 (2,0)&1.12 (4,0)&0.87 (3,2)&0.58 (5,2)&0.38 (5,3)&0.25 (7,4) \\
                 &&&                      0.32 (1,0)&2.28 (2,1)&0.52 (2,2)&0.46 (4,2)&0.49 (5,1)&0.26 (4,3)&0.23 (8,4) \\
                &&$\sigma_f$&0.0&0.0&0.18$\times 10^{-3}$&0.25$\times 10^{-3}$&0.85$\times 10^{-3}$&1.71$\times 10^{-3}$&2.48$\times 10^{-3}$ \\
\hline
$^{64}$Zn&14.06&$\sigma_a^T$&16.13 &12.84 & 8.57 & 6.16 & 4.68 &3.71 & 3.03\\
         &&$\sigma_r(\gamma,xnyp)$&9.67 (1,1)&4.67 (2,1)&3.19 (2,2)&1.64 (3,2)&0.82 (4,3)&0.54 (6,5)&0.43 (5,4) \\
                 &&&                    2.73 (3,2)&2.78 (1,2)&1.83 (4,3)&1.00 (5,3)&0.79 (5,4)&0.52 (5,3)&0.39 (6,5) \\
                 &&&                             2.08 (2,0)&1.42 (3,2)&1.03 (3,1)&0.67 (4,4)&0.50 (4,2)&0.40 (6,4)&0.32 (6,6) \\
                 &&$\sigma_f$&0.0&0.0&0.0&0.15$\times 10^{-3}$&0.0&0.28$\times 10^{-3}$&0.45$\times 10^{-3}$\\
\hline
$^{118}$Sn &21.19&$\sigma_a^T$ &19.32 &17.21 &12.84 &9.76 &7.67 &6.22 &5.16 \\
 &&$\sigma_r(\gamma,xnyp)$&12.95 (2,0)&15.03 (3,0)&9.44 (5,0)&5.74 (6,0)  &2.93 (8,0)&1.63 (9,0)&1.28 (9,1)\\
                &&&               6.33 (3,0)&1.91 (4,0)&2.57 (4,0) &2.12 (7,0)&2.03 (7,0)&1.57 (8,1)&0.97 (10,0)\\
                &&&                0.04 (1,1)&0.14 (2,1)&0.34 (4,1)&0.86 (5,1)&1.06 (7,1)&0.69 (8,0)&0.69 (10,0)\\
                 &&$\sigma_f$&0.0&0.0&0.0&0.0&0.0&0.0&0.0 \\
\hline
$^{140}$Ce &24.03&$\sigma_a^T$ &19.37 &18.00 &14.01 &10.88 &8.66 &7.08 &5.91 \\
 &&$\sigma_r(\gamma,xnyp)$&16.06 (2,0)&15.96 (3,0)&11.35 (5,0)&4.81 (6,0)&4.19 (8,0)&2.14 (9,0)&1.44 (10,1)\\
                &&&               3.23 (3,0)&1.66 (4,0)&1.48 (4,0)&3.94 (7,0)&1.45 (7,1)&1.43 (8,1)&1.03 (9,1)\\
                &&&              0.06 (1,1) &0.25 (2,1)& 0.58 (4,1)&1.04 (5,1)&1.15 (7,0)&0.94 (10,0)&0.72 (10,0)\\
                 &&$\sigma_f$&0.0&0.0&0.0&0.0&0.0&0.0&0.0 \\
\hline
$^{154}$Sm&24.96&$\sigma_a^T$&19.07 &18.20 &14.54 &11.44 &9.19 &7.55 & 6.32 \\
                 &&$\sigma_r(\gamma,xnyp)$&17.82 (3,0)&17.05 (4,0)&11.44 (6,0)&7.62 (8,0)&4.66 (10,0)&3.54 (11,0) &2.50 (12,0)\\
                 &&&              0.83 (2,0) & 0.84 (3,0)&1.61 (7,0)&2.04 (7,0)&3.05 (9,0)&1.45 (10,0)&0.84 (13,0)\\
                 &&&                 0.42 (4,0)&0.28 (5,0)&1.28 (5,0)&1.12 (9,0)&0.42 (10,2)&0.76 (12,0)&0.64 (11,1)\\
                 &&$\sigma_f$&0.0&0.0&0.0&0.0&0.0&0.0&0.0 \\
\hline
$^{174}$Yb&28.16&$\sigma_a^T$&18.67&18.49 &15.33 &12.29 &9.97&8.25 &6.95 \\
 &&$\sigma_r(\gamma,xnyp)$&18.02 (3,0)&17.38 (4,0)&11.72 (6,0)&5.79 (8,0)&5.65 (9,0)&3.27 (11,0)&2.92 (12,0)\\
                 &&&                     0.62 (2,0)&1.04 (3,0)&3.34 (5,0)&5.64 (7,0)&1.65 (8,0)&2.95 (10,0)&0.98 (11,0)\\
                 &&&                     0.03 (4,0)&0.05 (5,0)&0.08 (4,1)&0.30 (6,0)&1.64 (10,0)&0.44 (9,1)&0.91(13,0)\\
                 &&$\sigma_f$&0.0&0.0&0.0&0.0&0.0&0.62$\times 10^{-3}$&0.0 \\ 
\hline
$^{181}$Ta&29.44&$\sigma_a^T$&18.50&18.55 &15.58 &12.57 &10.24 & 8.49 & 7.17 \\              
 &&$\sigma_r(\gamma,xnyp)$&17.57 (3,0)&17.18 (4,0)&11.68 (6,0)&5.92 (7,0)&6.16 (9,0)&3.24 (11,0)&2.80 (12,0)\\
                 &&&           0.92 (2,0)&1.32 (3,0)&3.56 (5,0)&5.68 (8,0)&1.39 (10,0)&2.77 (10,0)&0.91 (11,1)\\
                 &&&           0.01 (1,1)&0.01 (4,2)&0.08 (4,1)&0.29 (6,1)&1.34 (8,0)&0.59 (9,1)&0.89 (11,0)\\
 &&$\sigma_f$&0.0&0.0&0.0&3.1$\times 10^{-3}$&1.02$\times 10^{-3}$&1.7$\times 10^{-3}$ &3.1$\times 10^{-3}$ \\
\hline
$^{197}$Au&31.68&$\sigma_a^T$ &17.97&18.55&16.03&13.12 &10.78 &8.99 &7.62 \\
 &&$\sigma_r(\gamma,xnyp)$&15.70 (3,0)&14.04 (4,0)&10.10 (6,0)&7.28 (7,0)&6.44 (9,0)&3.89 (10,0)&3.87 (12,0)\\
                 &&&       2.28 (2,0)&4.49 (3,0)&5.77 (5,0)&4.91 (8,0)&2.28 (8,0)&3.27 (11,0)&0.86 (11,0)\\
                 &&& 0.9$\times 10^{-3}$ (1,1)&0.01 (2,1)&0.06 (4,1)&0.57 (6,0) &1.25 (10,0)&0.40 (10,1)&0.80 (13,0)\\
                 &&$\sigma_f$&0.0&1.85$\times 10^{-3}$&9.22$\times 10^{-3}$&0.02&0.05&0.07&0.16 \\
\hline
$^{208}$Pb&32.33&$\sigma_a^T$ &17.48 &18.39&16.21&13.39 &11.07 &9.27 &7.88 \\
 &&$\sigma_r(\gamma,xnyp)$&15.47 (3,0)&16.12 (4,0)&11.79 (6,0)&6.49 (8,0)&6.97 (9,0)&5.12 (11,0)&3.53 (12,0)\\
                 &&&       2.01 (2,0)&2.23 (3,0)&4.27 (5,0)&6.46 (7,0)&2.41 (10,0)&2.61 (10,0)&2.16 (13,0)\\
     &&&            5.68$\times 10^{-3}$ (4,0)&0.01 (5,0)&0.03 (4,1)&0.19 (6,0)&1.18 (8,0)&0.53 (12,0)&0.77 (11,0)\\
                 &&$\sigma_f$&1.75$\times 10^{-3}$&0.02&0.06&0.12&0.20&0.33&0.44\\ 
\hline
$^{209}$Bi &32.96&$\sigma_a^T$ &17.49 &18.43 &16.28 &13.46 &11.13 &9.33 &7.93 \\
 &&$\sigma_r(\gamma,xnyp)$&15.71 (3,0)&15.91 (4,0)&10.90 (6,0)&6.65 (7,0)&6.60 (9,0)&3.97 (11,0)&2.40 (12,0)\\
                &&&               1.75 (2,0)&2.42 (3,0)&4.86 (5,0)&5.57 (8,0)&1.72 (10,0)&2.57 (10,0)&1.36 (13,0)\\
                &&&   6.12$\times 10^{-3}$ (2,1)&0.03 (2,1)&0.14 (4,1)&0.33 (6,1)&1.11 (8,0)&0.57 (10,1)&0.93 (11,1)\\
                 &&$\sigma_f$&0.02&0.06&0.21&0.40&0.60&0.88&1.18 \\ \hline
\hline
\end{tabular} 
\end{center}
\vspace{-0.55cm}
\end{table*}

\begin{figure}[htbp]
\vspace{0.0cm}
\eject\centerline{\epsfig{file=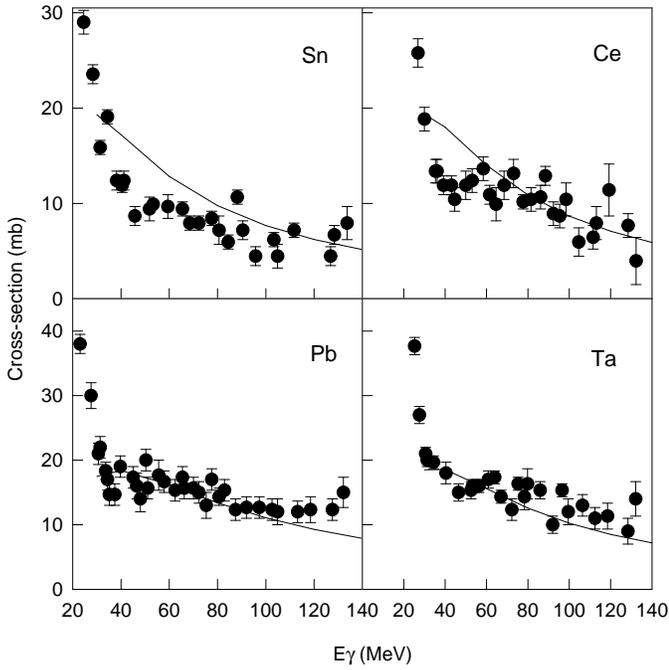,height=8.8cm,width=8.8cm}}
\caption
{Plots of total photoabsorption cross section versus photon energy. The continuous lines represent the total photoabsorption cross sections from the present calculation for discrete isotopes $^{118}$Sn, $^{140}$Ce, $^{181}$Ta and $^{208}$Pb whereas the data points are those corresponding to the measured values \cite{Le81} for natural targets of Sn, Ce, Ta and Pb.}
\label{fig2}
\vspace{0.0cm}
\end{figure}
\noindent

\begin{figure}[htbp]
\vspace{0.0cm}
\eject\centerline{\epsfig{file=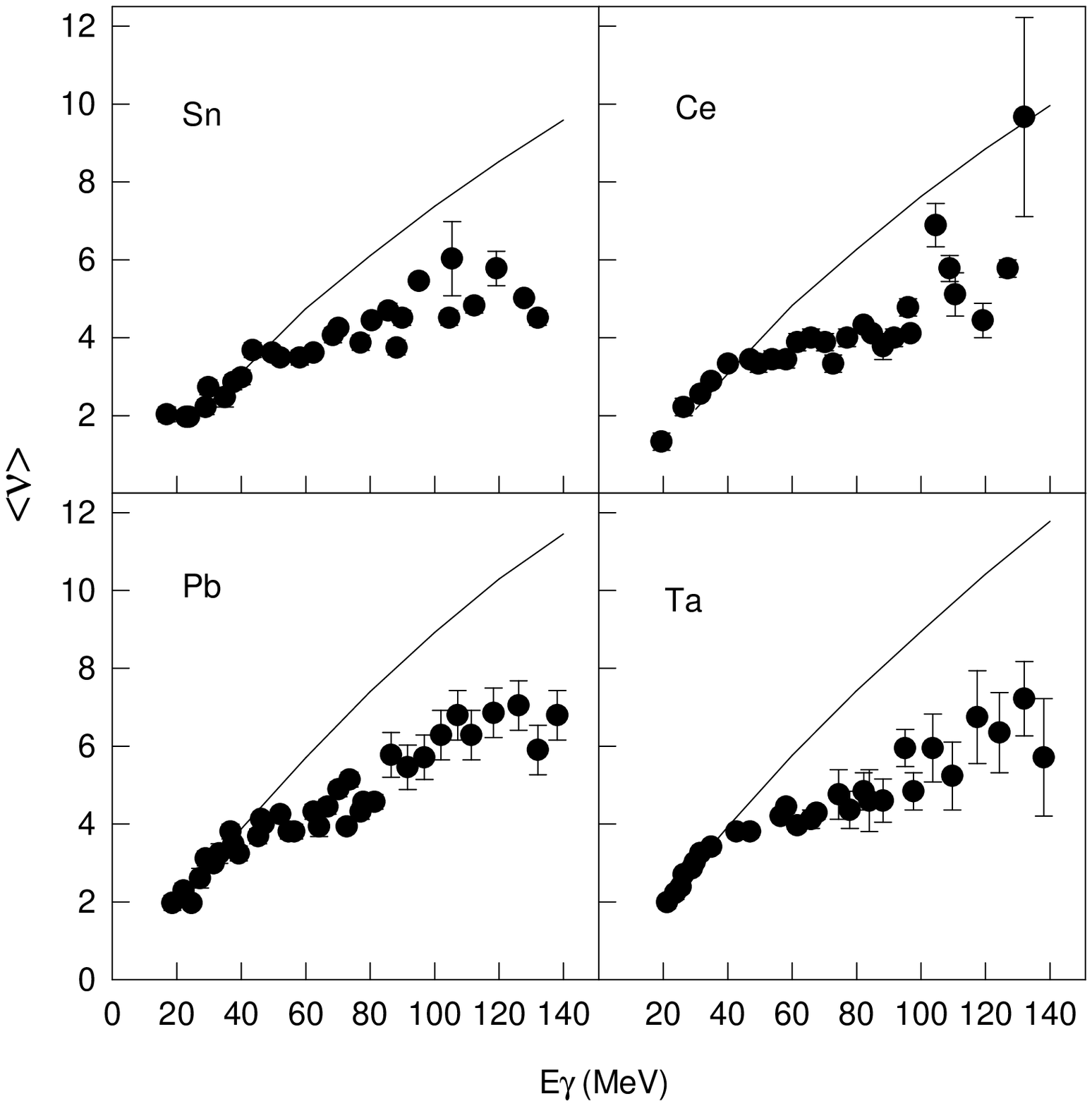,height=8.8cm,width=8.8cm}}
\caption
{Plots of average neutron multiplicities versus photon energy. The continuous lines represent the average neutron multiplicities from the present calculation for discrete isotopes $^{118}$Sn, $^{140}$Ce, $^{181}$Ta and $^{208}$Pb whereas the data points are those corresponding to the measured values \cite{Le82} for natural targets of Sn, Ce, Ta and Pb.}
\label{fig3}
\vspace{0.0cm}
\end{figure}
\noindent

    Each calculation is performed with 40000 events using a Monte-Carlo technique for the evaporation-fission calculation. This provides a reasonably good computational statistics. However, possible contributions from the direct reactions are omitted from the present calculations. The cross section $\sigma_r(E_\gamma)$ for any particular reaction channel $r$ is then calculated using the equation $\sigma_r(E_\gamma)= \sigma_a^T(E_\gamma) n_r/N$. Detailed calculations have been performed for $^{51}$V, $^{64}$Zn, $^{118}$Sn, $^{140}$Ce, $^{154}$Sm, $^{174}$Yb, $^{181}$Ta, $^{197}$Au, $^{208}$Pb and $^{209}$Bi at $E_\gamma=$ 30 MeV to 140 MeV. In Table I the results for ($\gamma,xn$) and ($\gamma,xnyp$) reaction cross sections are provided. The statistical error in the theoretical calculations for the photonuclear reaction cross sections can be estimated using the equation $\sigma_r \pm \Delta\sigma_r= \sigma_a^T [n_r \pm \sqrt{n_r}]/N$ which implies that  $\Delta\sigma_r=\sqrt{\sigma_a^T\sigma_r/N}$. In Fig.1 the plots of cross sections of evaporation residues resulting from ($\gamma,n$) and ($\gamma,np$) type of reactions for $^{64}$Zn for $E_\gamma=$ 30 MeV and 40 MeV are shown as typical cases. In Fig.2 the plots of total photoabsorption cross section versus photon energy $E_\gamma$ are presented. The continuous lines represent the total photoabsorption cross sections for Sn, Ce, Ta and Pb from the present calculations whereas the points with error bars are the experimental data \cite{Le81}. Fig.3 shows the plots of average neutron multiplicities $<\nu>$=$\Sigma_x x\sigma(\gamma,xnyp)/\sigma_a^T$ versus photon energy $E_\gamma$. The continuous lines represent the average neutron multiplicities for Sn, Ce, Ta and Pb from the present calculations whereas the points with error bars are the experimental data \cite{Le82}. Considering the fact that the present calculations for photoabsorption cross sections are for discrete isotopes $^{118}$Sn, $^{140}$Ce, $^{181}$Ta and $^{208}$Pb whereas the measurements \cite{Le81} were performed using natural targets of Sn, Ce, Ta and Pb, present results agree quite closely with the available experimental data at intermediate energies. This effect is more pronounced in  the plots for $<\nu>$ where at $E_\gamma$ above 50 MeV, present calculations gradually overestimate. This behaviour is expected as neutron yields would be less in presence of lower mass isotopes in natural targets. Although the present simple model appears to start breaking down above photon energies 50-60 MeV, more experimental data are required before any definitive conclusion can be reached. It is worthwhile to mention here that the present formalism provides excellent estimates \cite{Mu07} for photofission cross sections of actinide and pre-actinide nuclei at intermediate energies. 

    The agreement of the present theoretical calculations according to the QD model with the experimental data for photoabsorption cross sections $\sigma_a^T(E_\gamma)$ is much better than that for the average neutron multiplicities $<\nu>$ which can be visualized from Fig.2 and Fig.3 respectively. Therefore, the main physics issue remains in the second stage of the process. In an earlier work, good agreement for average neutron multiplicities $<\nu>$ were achieved by inclusion of the precompound decay process \cite{Bl83} to account for higher neutron yield than estimated by the evaporation model alone. Certainly, similar modification can only worsen the present results. The implication of such a state of affairs suggests, perhaps, the need for more precise measurements since even one or two missing neutrons in the measurements can account for the discrepancy in $<\nu>$. This point is further strengthened by a recent work \cite{Ps05} where statistical decay of the excited compound nucleus was simulated in a  Monte-Carlo framework like the present work \cite{Mu07} to describe the evaporation process. Both provided excellent estimates for photofission cross sections of actinide nuclei in a wide range of energies ($E_\gamma=$ 68 MeV - 3.77 GeV), ranging from the QD region ($\sim 30-140$ MeV) \cite{Mu07} to the region of intranuclear cascade model (above the pion production threshold) \cite{Ps05}.

    In summary, the cross sections for the fission and the evaporation residues are calculated for photon induced nuclear reactions at intermediate energies. Monte-Carlo calculations for the evaporation-fission competition are performed assuming 40000 incident photons for each calculation which provides a reasonably good statistics for computationally stable results. Present calculations provide good estimates of cross sections for the reaction channels ($\gamma,xn$) and ($\gamma,xnyp$) for nuclei $^{51}$V, $^{64}$Zn, $^{118}$Sn, $^{140}$Ce, $^{154}$Sm, $^{174}$Yb, $^{181}$Ta, $^{197}$Au, $^{208}$Pb and $^{209}$Bi at $E_\gamma=$ 30 MeV to 140 MeV. No fission event has been observed below 40 MeV for these nuclei except few events in cases of $^{209}$Bi and $^{208}$Pb. Additional experiments may be necessary for more precise measurements of neutron multiplicities.

\end{document}